\documentclass[11pt]{article}
\usepackage{amsmath,amssymb,geometry,booktabs,url}
\title{A Geometry-Grounded Data Perimeter in Azure}

\author{
        Christophe Parisel\thanks{Email: ch.parisel@gmail.com}
}
\date{}

\begin{document}
\maketitle

\begin{abstract}
	While \emph{Data Perimeter} is ubiquitous in cybersecurity speak, it rarely defines how boundary points are arranged. In this paper we show how Azure’s blast radius ultrametric provides the distance, and how solving the Traveling Salesman Problem in this ultrametric space provides the ordering, yielding a true geometric contour: an actionable perimeter measure for Service Principals (SPNs) data plane operations de-escalation.
\end{abstract}

\section{Introduction}
In cybersecurity and data privacy management, the notion of a Data Perimeter is rarely tied to a concrete geometric construction, because perimeter implies a \emph{distance} function and a \emph{circular ordering} of the data, both of which are usually either missing or don't make sense for abstract datasets.
To define a meaningful Data Perimeter in Azure, we leverage the \emph{blast radius} ultrametric\cite{br} to propose a geometry-grounded perimeter which is both efficiently computable and minimal.

\medskip
\noindent

We show how our Data Perimeter acts as a secondary sort key after blast radius stratification, to provide fine-grained risk prioritization of SPNs.

\section{Background}
\subsection{Related Work}
\subsubsection{Attack Surface Model}
Manadhata and Wing’s attack surface\cite{surface} pioneered the formalization of "entry/exit points" and "channels" as quantitative risk contributors. While their model provides a useful abstraction for OS-level systems, it is less suited to the structured permission sets typical of cloud platforms IAM like Azure RBAC.

In contrast, our approach treats the action set geometry directly, producing shape-aware summaries rather than quantitative counts.

\subsubsection{Ultrametrics and Hierarchical Models}
Ultrametric spaces arise naturally in settings when items are grouped by resource hierarchy. Such metrics obey the strong triangle inequality and induce a natural tree structure. Previous works on hierarchical clustering have used ultrametrics to model efficient machine learning algorithms\cite{ml}, phylogenetic trees\cite{philogen}, and linguistic evolution\cite{linguistics}.

Our use of ultrametrics follows this tradition, but we leverage their properties for efficient tour construction (not just visualization).

\subsubsection{Seriation and metric TSP}
Seriation, the process of ordering elements to reveal latent structure, has applications in archaeology, heatmap visualization, and gene expression analysis. In computational geometry, finding an optimal ordering that minimizes some distance (e.g., via the Traveling Salesman Problem) has been proposed as a generic sorting principle\cite{seriation}. 

We contribute a novel application of this idea to the Cloud context, in Azure RBAC, showing that TSP solutions over ultrametrics are not only efficient to compute but also yield meaningful secondary metrics.

\subsubsection{Perimeter-Like Measures in Security}
Few existing metrics capture the notion of a perimeter over privileges. Graph-based approaches like privilege escalation trees (e.g., BloodHound\cite{blood}) focus on path enumeration and control flow, not data-action layout. Similarly, role-mining algorithms often compress permissions but ignore spatial structure. 

Our notion of a Data Perimeter is grounded in metric space geometry, enabling fine-grained differentiation within otherwise identical blast-radius bands.

\subsection{SPNs and Data Actions in Azure}
Service Principals in Azure hold sets of fine-grained data actions (e.g., ReadBlob, WriteSecret) scoped to individual resources or to resource containers (resource groups, subscriptions, management groups). Ultimately, this design means that Azure Data Perimeter is not a global, Tenant-wide metric: it depends on each SPN.

\medskip
\noindent

SPN permissions can be assigned either directly, or via groups membership. Excessive read permission may lead to a data leakage, whereas excessive 
write permissions may lead to data forgery. The concentration of read and write permissions across a large set of resources or resource containers under the same SPN makes risks assessment related to lateral motion particularly challenging.

\subsection{Blast Radius Ultrametric}

The Blast radius was introduced to address the challenge of measuring lateral motion, data leakage and data forgery of individual SPNs in Azure. It is fully implemented in Azure Silhouette\cite{silhouette}.

\medskip
\noindent

Define an ultrametric distance $d(a,b)=\tfrac{\mathrm{impact}(a,b)}{2^{2D(a,b)+1}}$, where $D$ is the Least Common Ancestor of $a$ and $b$ in Azure's native clustering hierarchy, and $impact$ is a parameter depending on data actions\cite{br}. 

\medskip
\noindent

As with all ultrametric distances, it satisfies the strong triangle inequality:

\[
\forall x, y, z \in X, \quad d(x, z) \leq \max\{d(x, y), d(y, z)\}
\]

\noindent

The \emph{blast radius} of an SPN is the diameter of $d()$: $\max_{i,j}d(x_i,x_j)$. It ranges from 0.0 (no permissions) to 1.0 (Tenant-wide permissions).

\newpage

\section{Problem Statement}
The Blast Radius assigned to each SPN is a sort key to quickly identify the most risky SPNs.  Due to the nature of ultrametry and the use of the $impact$ parameter, SPNs are grouped into up to 22 bands of identical Blast radius values:
\begin{itemize}
	\item 2 bands for Tenant-wide radius
	\item 12 bands for Management groups
	\item 2 bands for subscriptions
	\item 2 bands for resource groups
	\item 2 bands for resources
	\item 2 bands for resource parts
\end{itemize}

The Blast radius stratifies SPNs into coarse bands, so many SPNs tie on the same bands, leaving no intra-band ranking. 

\medskip
\noindent

We seek a secondary sorting key that leverages the ultrametric structure to distinguish SPNs further.

\section{Data Perimeter}
\subsection{Formulation}
Given data actions ${x_1,\dots,x_n}$ with ultrametric $d$, the Data Perimeter is defined as the minimal circular path covering all actions under ultrametric distance. 

\[
P = \min_{\pi \in S_n} \sum_{i=1}^{n} d(x_{\pi(i)}, x_{\pi(i+1)}), \quad \text{with } x_{\pi(n+1)} = x_{\pi(1)}
\]

Where $\pi$ is a permutation of the data action indices. 

\medskip
\noindent

This amounts to solving a metric Traveling Salesman Problem (metric TSP). 
A major benefit of using the ultrametric TSP formulation for computing the Data Perimeter is that the problem, typically NP-hard in general, becomes efficiently solvable. In ultrametric spaces, the hierarchical structure imposed by the strong triangle inequality enables polynomial-time solutions to the metric TSP. This stems from the fact that shortest tours respect the natural tree structure of the space. That is, the tour effectively follows a traversal of the leaves of the ultrametric tree with minimal backtracking.

\medskip
\noindent

As shown in the ultrametric TSP pseudo-code implementing the nearest-neighbor algorithm below, the Data Perimeter is not only geometrically grounded but also computationally tractable in our setting. 

\newpage

\subsubsection{Ultrametric TSP pseudo-code}

\textbf{Input:} A set of data actions $A = \{a_1, \dots, a_n\}$ with distance function $d(a, b)$

\noindent
\textbf{Output:} Perimeter length $L$

\begin{enumerate}
  \item Let $P \leftarrow [\,]$ \hfill (Tour list)
  \item Pick $a_{\text{start}} \in A$ at random
  \item Append $a_{\text{start}}$ to $P$
  \item Let $U \leftarrow A \setminus \{a_{\text{start}}\}$ \hfill (Unvisited actions)
  \item \textbf{while} $U \neq \emptyset$ \textbf{do}
  \begin{enumerate}
    \item Let $a_{\text{current}} \leftarrow$ last element of $P$
    \item Find $a_{\text{next}} \in U$ such that $d(a_{\text{current}}, a_{\text{next}})$ is minimized
    \item Append $a_{\text{next}}$ to $P$
    \item Remove $a_{\text{next}}$ from $U$
  \end{enumerate}
  \item Append $a_{\text{start}}$ to $P$ to complete the cycle
  \item Initialize $L \leftarrow 0$
  \item \textbf{for} $i = 1$ to $|P| - 1$ \textbf{do}
  \begin{enumerate}
    \item $L \leftarrow L + d(P_i, P_{i+1})$
  \end{enumerate}
  \item \textbf{return} $L$
\end{enumerate}

\subsubsection{Notes}
\begin{itemize}
        \item Ultrametric TSP admits multiple optimal tours, so the circular ordering required for constructing a perimeter is not unique, but all tour lengths are equal. They form an equivalence class.
        \item Because of the wraparound, a tour length doesn't depend on its starting point.
\end{itemize}

\subsubsection{Known Limitations}

Silhouette implements Blast radii calculation over the native Azure Tenant hierarchy, as well as on a family of
alternate hierarchies. This option lets risk officers tailor blast radii to the local needs of their organization: they can for instance
use a special alternate hierarchy to represent the state of a reorganization, a merger or acquisition, a carve-out, ...

\medskip
\noindent

When hierarchical families are defined, Silhouette computes the Blast radius as the pointwise infimum in all hierarchies to yield a tighter, smaller
Blast Radius than the native one. In doing so, it must be noted that the pointwise infimum is not ultrametric, hence the data Perimeter should be treated with care in this situation. Falling back to the native tenant hierarchy is highly recommended.

\medskip
\noindent
\newpage

Here is a small counter-example showing the lack of ultrametry:

Let \( X = \{x, y, z\} \), and define two ultrametrics \( d_1 \) and \( d_2 \) on \( X \) as follows:

\[
\begin{aligned}
&d_1(x, y) = 2,\quad d_1(y, z) = 1,\quad d_1(x, z) = 2, \\
&d_2(x, y) = 1,\quad d_2(y, z) = 2,\quad d_2(x, z) = 2.
\end{aligned}
\]

Each \( d_i \) satisfies the ultrametric inequality:
\[
d_i(a, c) \le \max\{d_i(a, b), d_i(b, c)\} \quad \text{for all } a,b,c \in X.
\]

Now define \( d_3 \) as the pointwise minimum:
\[
d_3(a, b) := \min\{d_1(a, b), d_2(a, b)\}.
\]

This gives:
\[
\begin{aligned}
&d_3(x, y) = \min\{2, 1\} = 1, \\
&d_3(y, z) = \min\{1, 2\} = 1, \\
&d_3(x, z) = \min\{2, 2\} = 2.
\end{aligned}
\]

However, the strong triangle inequality fails:
\[
d_3(x, z) = 2 \not\le \max\{d_3(x, y), d_3(y, z)\} = \max\{1, 1\} = 1.
\]

\section{Mean distance ($\mu_{\mathrm{mean}}$)}
The average distance between any two data actions is the metric that springs to mind for complementing the Blast radius. Unfortunately, if it captures global dispersion, it ignores local structure and chaining. In this section, we explain why this doesn't make it a good secondary key candidate for sorting SPNs.

\subsection{Comparison with Data Perimeter}
To understand the benefits of the Data Perimeter over the mean, we sample 1285 SPNs that we sort by Blast band: 
\begin{enumerate}
	\item For each SPN we compute a \emph{spread\_ratio} defined as $\frac{\text{P}}{n \cdot \mu_{\mathrm{mean}}}$, where $n$ is the data actions count held by this SPN. The perimeter $P$ is normalized by the action count $n$ to maintain dimensional consistency with $\mu_\mathrm{mean}$.
	\item We group SPNs by band 
	\item We calculate the average spread ratio in each band.
\end{enumerate}

\medskip
\noindent

Table 1 summarizes our findings. Blast radii are first shuffled, then anonymized into Band IDs. 

\begin{table}[ht]
\centering
\begin{tabular}{cccc}
\toprule
	Band ID & SPNs count & Spread Ratio & Regime\\
\midrule
	I & 33   & 1.00000 & Tight\\
	II & 69   & 0.998993 & Tight\\
	III & 4   & 0.820245 & Dispersed\\
	IV & 2   & 1.00000 & Tight\\
	V & 110   & 0.99186 & Tight\\
	VI & 8   & 0.978792 & Dispersed\\
	VII & 228   & 1.00000 & Tight\\
	VIII & 19   & 1.00000 & Tight\\
	IX & 3   & 1.00000 & Tight\\
	X & 505   & 0.997459 & Tight\\
	XI & 6   & 1.00000 & Tight\\
	XII & 70   & 0.987850 & Tight\\
	XIII & 2   & 1.00000 & Tight\\
	XIV & 16   & 1.00000 & Tight\\
	XV & 28   & 0.581303 & Dispersed\\
	XVI & 66   & 0.806384 & Dispersed\\
	XVII & 2   & 1.00000 & Tight\\
	XVIII & 75   & 0.950056 & Dispersed\\
	XIX & 5   & 0.956395 & Dispersed\\
	XX & 2   & 1.00000 & Dispersed\\
	XXI & 32   & 0.888986 & Dispersed\\

\bottomrule
\end{tabular}
\caption{In-band spread ratios (Data Perimeter vs. mean) per shuffled and anonymized blast radius band.}
\label{tab:spread}
\end{table}

\subsubsection{Discussion}

The in-band \texttt{spread\_ratio} values (Table~\ref{tab:spread}) fall into two distinct regimes:

\paragraph{Tightly clustered SPNs (blast radius $\lesssim10^{-4}$)}  
In bands VII-XIV, and other small radius groups, the normalized Data Perimeter and the mean coincide (spread\_ratio~$\approx1.0$). Here, each SPN’s permission set is nearly point-like, so any reasonable ordering of actions (whether implicit in the mean or made explicit via TSP) produces the same minimal contour length.

\paragraph{Dispersed SPNs (blast radius $\gtrsim10^{-4}$)}  
As blast radii grow, we see a systematic divergence (spread\_ratio~$<1$). For example:
\begin{itemize}
  \item Band XV: spread\_ratio~0.58  
  \item Band XVI: spread\_ratio~0.81  
  \item Band XVIII: spread\_ratio~0.95  
\end{itemize}
This reflects the fundamental difference between our two metrics.
\begin{itemize}
  \item The \emph{mean} sums large, disjoint jumps and thus overestimates overall spread.
  \item The \emph{TSP-based perimeter} constructs a continuous tour, chaining cluster centers and “shortcutting” around outliers. By exploiting the strong triangle inequality, it compresses lengthy detours into shorter, hierarchical paths.
\end{itemize}

\paragraph{Implications for SPN prioritization}  
When many SPNs share the same blast radius, the Data Perimeter provides a fine-grained, geometry-aware tiebreaker. It surfaces principals whose permission sets exhibit branching or elongated ultrametric structures whereas the mean alone would treat them identically to more compact SPNs.

\medskip
\noindent

This richer signal enables security teams to target reviews and policy adjustments not just based on “how far” an SPN can reach, but on “how” its permissions are organized.  It confirms that TSP-derived perimeter is not just another average: it captures the global connectivity structure of data actions.

\section{Conclusion}
We introduced the Data Perimeter as a geometry-grounded refinement of the blast radius, capturing the spatial layout and redundancy of data 
actions assigned to service principals (SPNs). Unlike scalar averages such as the ultrametric mean, the perimeter is sensitive to seriation, clustering, and the underlying topology of permission sets.

\medskip
\noindent

By leveraging ultrametric distance and a tractable TSP solver, we demonstrated that the Data Perimeter can be computed efficiently and yields a non-arbitrary circular ordering of actions. This ordering is meaningful, not merely heuristic: it reflects a class of equivalent perimeters that respect the hierarchical structure of permissions, compressing global dispersion into a single tour.

\medskip
\noindent

Our small comparative analysis across blast radius bands revealed that while ultrametric mean and perimeter agree in tightly scoped SPNs, they diverge significantly as permission sets grow complex. The Data Perimeter remains stable and informative even in the presence of branching or fragmented access patterns, highlighting its suitability as a secondary sorting key for SPNs sharing similar blast magnitude.

\medskip
\noindent

This work bridges abstract privilege metrics with actionable geometric intuition. It enables analysts to go beyond “how much power” toward “how this power is organized,” providing a new lens to prioritize, visualize, and reduce risk in large-scale cloud environments.

\bibliographystyle{plain}

\section*{Appendix: Minimal Data Perimeter}

In this work, we have argued that the Data Perimeter in Azure security models behaves not just as a metaphor, but as a measurable geometric quantity. To further give substance to this geometric view, we identify the shapes corresponding to the smallest Data Perimeter.

\subsection*{Ultracycle and Minimal Perimeter}

Given an ultrametric space $(X, d)$ with $n$ points, an \emph{ultracycle} is a configuration where all pairwise distances are equal: $d(x_i, x_j) = \xi$ for all distinct $i, j \in \{1, 2, \ldots, n\}$ and some constant $\xi > 0$.

In any ultrametric space, ultracycles uniquely minimize the TSP tour length among all configurations with the same number of points.

\subsubsection*{Proof}
Let $(X, d)$ be an ultrametric space and consider $n$ points $\{x_1, x_2, \ldots, x_n\} \subset X$.

\medskip

\textbf{Case 1: Ultracycle Configuration}

Suppose all pairwise distances are equal: $d(x_i, x_j) = \xi$ for all $i \neq j$.

For any TSP tour $\pi$ visiting all points exactly once, the tour length is:
\[
	P_{\text{ultra}} = \sum_{i=1}^{n} d(x_{\pi(i)}, x_{\pi(i+1)}) = \sum_{i=1}^{n} \xi = n\xi, \text{   where } x_{\pi(n+1)} := x_{\pi(1)}.
\]

\medskip

Every possible tour has identical length $n\xi$.

\medskip

\textbf{Case 2: Non-Ultracycle Configuration}

Now suppose the configuration is not an ultracycle. Then there exist distinct indices $i, j, k, \ell$ such that:
\[
d(x_i, x_j) \neq d(x_k, x_\ell)
\]

\medskip

Let $d_{\min} = \min_{i \neq j} d(x_i, x_j)$. 

\medskip

\textbf{Claim:} Any TSP tour must include at least one edge with distance $> d_{\min}$.

\medskip

\textbf{Proof of Claim:} 

Suppose, for contradiction, that there exists a tour using only edges of distance $d_{\min}$. 

\medskip

For any three points $x_i$, $x_j$, $x_k$:
\begin{itemize}
	\item If $d(x_i,x_j) = d_{min}$ and $d(x_j,x_k) = d_{min}$, then $d(x_i,x_k) \leq \max\{d(x_i,x_j), d(x_j,x_k)\} = d_{min}$ 
	\item Since $d_{min}$ is the global minimum distance, $d(x_i,x_k) \geq d_{min}$
	\item Thus, $d(x_i,x_k) = d_{min}$
\end{itemize}

By induction, any two points connected by a path of $d_{min}$ edges must be at distance $d_{min}$.

Since the tour connects all $n$ points via $d_{min}$ edges, all pairwise distances must be $d_{min}$, forming and ultracycle.

This contradicts the non-ultracycle assumption.

\medskip

Ultracycles uniquely minimize the TSP tour length in ultrametric spaces.

\subsection*{Security Implications}

In Azure's permission model, a Principal with $n$ data actions all scoped to the same resource (forming an ultracycle at depth $d$ in the hierarchy) achieves the minimal Data Perimeter, representing optimal privilege containment.

This geometric result aligns with the theoretical foundation for the least-privilege principle: permissions should be clustered as tightly as possible in the resource hierarchy to minimize potential lateral movement exposure.

\end{document}